# Dynamics of competing species in a model of adaptive radiations and macroevolution

Birgitte Freiesleben De Blasio[1], Fabio Vittorio De Blasio[2],

[1] Department of Statistics, Institute of Basic Medical Sciences, University of Oslo, Norway,

[2] Department of Geosciences, University of Oslo, Norway.

Corresponding author:

Dr. Fabio Vittorio De Blasio
c/o International centre for Geohazards
P.O. Box 3930 Ullevaal Stadion
N-0806 Oslo
Norway

Phone: +47 22 02 31 22
E-mail: fvb@ngi.no

## *ABSTRACT*

We present a simple model of adaptive radiations in evolution based on species competition. Competition is found to promote species divergence and branching, and to dampen the net species production. In the model simulations, high taxonomic diversification and branching take place during the beginning of the radiation. The results show striking similarities with empirical data and highlight the mechanism of competition as an important driving factor for accelerated evolutionary transformation.

## *INTRODUCTION*

The process of adaptive radiation (AR) is believed to play a major role in the evolution of diverse life forms on Earth. In the fossil record, large-scale AR's are seen as 'explosions' of new taxonomical groups during particularly active periods in time. The greatest AR of all times was the Cambrian explosion that gave rise to most of all known animal phyla. Small-scale AR's usually occur in limited and isolated environments. A prime example is Darwin's finches on the Galapagos Island that have adapted from a single immigrating species to occupy different ecological niches [1].

Despite great variability in size, AR's share common characteristics. (1) They are initiated when new resources becomes available to a founder species, e.g. because it develops a key character, or in the wake of a mass extinction. (2) The availability of resources triggers rapid evolution of morphologically distinct groups to fill ecological niches not yet occupied. (3) The creative phase is followed by a longer phase of species multiplication with limited creation of novelty [2].

A characteristic pattern of AR's is that high taxonomic groups are established early and anticipate the creation of low level groups (Fig. 1). Why does the branching of major groups take place at the origin of the radiation and becomes rare afterwards?

The question is part of a fundamental controversy that spans the last century of evolutionary thinking, namely if the origin and proliferation of novelties has been gradual—or—if history of life has moved in leaps. In other words, are the forces that shape long-term evolution (macroevolution) identical to the ones that operate at the level of individuals (microevolution)?

According to the widely accepted Neo-Darwinian Theory [2,3], adaptive evolution is a gradual process that is well explained by the conjunction of mutation, natural selection and genetic drift. Still, other evolutionists argue that macro-evolutionary patterns are more than simply the accumulation of micro-evolutionary processes over long periods of time. For instance, in the macro-mutation hypothesis [4] it is claimed that major taxonomic groups have formed as a result of infrequent, but large 'jumps' in genotypic appearance. More recently, it is stated in the Theory of Punctuated Equilibrium [5] that evolution proceeds through long periods of stasis 'punctuated' by rapid bursts of speciation. The fast proliferation, in turn, leads to selection on higher taxonomic levels and introduces a hierarchical and directed structure to macroevolution [6].

Neither theory is able to account for the uneven appearance of high-level taxa in the fossil record. For example, if large morphological jumps may occur at any time, why is the creation of high taxonomical categories only present in the beginning of a radiation? And, which forces prevent the radiation from progressing forever?

It has been argued that genetic stability has increased during evolution, thereby favouring early development of higher taxa [7]; the hypothesis of genetic robustness has also been studied theoretically [8]. An alternative hypothesis that has gained increasing acceptance is that major morphological innovation decreases during the radiation as a consequence of ecological saturation following niche occupation [9]. Valentine et al. [10,11] have used computer models to study the influence of eco-space colonisation on AR's. In the models, an ecological niche capable of supporting a species is represented by a tessera in a two-dimensional 'adaptive space'. Thus, each tessera can be occupied by at most one species, and species may colonise new tesserae by sending a daughter species. Local colonisation of a neighbouring cell (microevolution) is assumed to be more frequent than large jumps spanning the distance of more tessera (macroevolution). Large jumps are mostly successful in the beginning of the radiation when space is relatively free, and hence, the model rightfully explains the uneven appearance of major groups. However, the concept of macro-mutation is controversial, and the use of tessera makes the model stiff.

Competitive interactions play a key role in ecological dynamics, and competition also appears to be essential for species proliferation. Noticeably, an AR often produces morphologically different species where seemingly no adaptive reason for the diversification exists; e.g. the finches on the Galapagos Islands are similar in key characteristics when occurring on different islands, while they differ markedly when present on the same island. The non-adaptive diversification shows that competition between species exploiting the same resources is important for creation of diversity [1] Recent laboratory bacterial experiments have significantly increased our understanding of evolutionary processes and suggest that species branching is promoted by competition [12,13]. The test-tube bacterial experiments show spatial patterns, which resemble grand-scale AR's observed in fossil lineages. This apparent scale-independence may reflect that small- and large-scale AR's are driven by the same forces of mutation and natural selection [14].

We present a simple model for the dynamics of AR's based on inter-species competition. In the model we use a morphospace representation of species, which measures the disparity of organisms in shape, form and structure. Each axis quantifies a single phenotypic character, and each species is represented by a single point in the space when measured for the various characters. Phenotypic similarity between species implies that they are likely to exploit the same resources [15]. The observation is used to 'translate' competition among species for ecological niches and resources into strong competition among neighbouring species in morphospace. However, even species that are far morphologically may compete for common resources, like the competition for light among plants. Long-range competition is accounted for by allowing a finite tail in the competitive interaction for morphologically distant species. The morphometric framework allows a direct comparison with the rate of change and variability in forms in the fossil record.

Methods borrowed from complex system analysis and statistical physics have been applied to study the extinction and origination statistics in the fossil record [16-18]. Instead, only few models have addressed the creative phase of evolution. In the present work we study the dynamics of adaptive radiations by seeking for significant morphospace occupation patterns based on a set of simple rules. The model demonstrates that during periods of plentiful ecological opportunities, it is possible to produce rapid and extreme diversification through gradual changes in morphology.

## THE MODEL

Let each species $S$ be characterised by a fixed number $N$ of phenotypes $P = \{p_1, p_2, \ldots p_N\}$, and let the different phenotypic traits be independent. Then we can symbolise each species in an $N$-dimensional morphospace $M_N$ by a single point. We will use $N = 2$ as a reasonable compromise between unrestricted freedom in the dynamics and fast computation and possibility of visualisation. The model dynamics is as follows:

1. ***Initialisation:*** At the beginning of each simulation, one single species is dumped at a central position in $M_N$.

**2. *Speciation:*** New species are formed at each time-step $\Delta t$ with a speciation rate $s$, which is constant for all species. New species originates as a propagule from the parent species in a random direction; the distance between parent-offspring is taken to be Gaussian distributed with variance $\sigma s$.

**3. *Competition:*** Species compete for resources and may become extinct if these resources are shared among too many competitors. The extinction probability per time unit $p_{ext}$ of species $S_i$ depends on density and relative position on the other species, whilst it is independent on absolute position in $M_N$. It is calculated as a sum over all species of the competition function $f$:

$$p_{xt}(S_i) = \varepsilon \sum_{j \neq i} f(d_{i,j}); \qquad d_{i,j} = \sqrt{(x_i - x_j)^2 + (y_i - y_j)^2} \qquad (1)$$

where $d_{i,j}$ is the distance between species $S_i$ and species $S_j$, and $\varepsilon$ is a constant. Lacking a sound biological argument for the functional form of $f$, different functions (I-II) were studied and the robustness of the results were tested against these forms.

**Model I: *The smoothed sphere:***

$$f_I\left(d_{i,j}\right) = \frac{f_0 - f_\infty}{1 + \exp\left(\dfrac{d_{i,j} - R}{a}\right)} + f_\infty \qquad (2)$$

where the radius of short-range competition is controlled by an effective radius $R$, and the smoothness of influence is quantified by the diffusivity $a$. The competitive interaction between two species is $f \approx f_0$ (short-range) when their distance $d_{i,j} < R$ and becomes $f \approx f_\infty$ when $d_{i,j} > R + 2a$. In particular, for $a \to 0$ the function becomes step-sized (hard sphere).

**Model II: A *step function with power-law tail***

$$\begin{aligned} f_{II}\left(d_{i,j}\right) &= f_0 &&\text{for } d_{i,j,} < R \\ f_{II}\left(d_{i,j}\right) &= f_0\left(\frac{d_{i,j}}{R_L}\right)^{-\beta} &&\text{for } d_{i,j} \geq R \end{aligned} \qquad (3)$$

where again $R$ is the radius of short-range competition, $R_L$ is the long-range competition range and $\beta$ is the scaling constant of the power-law tail.

The fitness concept is here based on the morphological disparity of a species in comparison with other species, rather than being defined as intrinsic morphological advantages. Hence, in a strict sense the model dynamics does not contain any real 'adaptation' and the term adaptive radiation is therefore intended in a broad sense. However, introducing a space-dependent fitness would complicate the model without providing new insight to the main study of competitive interactions of species on morphospace. The simulations are run for up to 300,000 time-steps. The basic ingredients of the model are summarized in Fig. 2.

## *RESULTS*

Numerical calculations were performed with different forms of the short- and long-range competition function. In all simulations the short-range radii $R$ was chosen larger than the speciation range $\sigma$ so that speciation takes place within the range of the short-range competition.

## **MORPHOLOGICAL DIVERSIFICATION**

Fig. 3 shows the results of a simulation with a power-law decaying tail of the competition function (model II). The number of species follows a growth curve (Fig. 3A) with rapid diversification in the beginning that slowly decelerated to reach a constant level. A closer look at the early phase (insert Fig. 3A) reveals that the increase in species number is step-wise interrupted by static periods that prolongs with time. The non-uniform growth in the beginning of the radiation is further illustrated by plotting the mean square morphological distance of the entire species distribution (Fig. 3B). It is seen that the separation in phenotypic traits after a sufficiently long time increases roughly with $\sqrt{t}$, indicating a diffusion-controlled divergence in morphological appearance of the species. The effect is independent on the form of competition function, provided it does not drop abruptly as a function of the distance in morphospace.

The general features of the model can be understood analytically by studying the species density distribution $\Lambda(\vec{x},t)$. When $R > \sigma$, the species development can be treated as a point-like process, and $\Lambda(\vec{x},t)$ evolves approximately as

$$\frac{\partial \Lambda(\vec{x},t)}{\partial t} \approx s\Lambda(\vec{x},t) - \hat{E}\Lambda(\vec{x},t) + K\sigma^2 s\nabla^2\Lambda(\vec{x},t) \qquad (4)$$

where $s$ is the speciation rate and $K$ is a geometrical constant. The first term defines local proliferation of species, the second term is the extinction caused by competition, and the last term is a diffusion term accounting for the fact that speciation is non-local with a range $\sigma$. The extinction kernel is given by

$$\hat{E} = \varepsilon \int_{M_N} d^2x\, f(\vec{x} - \vec{x}')\Lambda(\vec{x}',t) \qquad (5)$$

that is calculated from integration over the entire morphospace. First, we consider the long-distance behaviour $(D \gg R)$; substituting the competition function $f$ with its average value $\bar{f}$, the extinction kernel can be written as $\hat{E} \approx \varepsilon \bar{f} N(t)$, where $N(t)$ is the total number of species

$$N(t) = \int_{M_N} d^2x\, \Lambda(\vec{x},t) \qquad (6)$$

By integrating Eq. (6) and using Eq. (4), we find upon application of Gauss' theorem the time-differentiated species number to be

$$\frac{dN(t)}{dt} \approx s\,N(t) - \varepsilon \bar{f}\, N^2(t) \qquad (7)$$

which shows that the growth is sigmoidal, i.e. the species number initially increases exponentially to reach an asymptotic value $N(t \to \infty) \equiv \bar{N} = s/(\varepsilon \bar{f})$. The limit corresponds to the plateau in Fig. 3A.

We now turn to the large-scale behaviour of the model. For simplicity we may neglect short-range competition and the evolution of the species density distribution becomes

$$\frac{\partial \Lambda(\vec{x},t)}{\partial t} = s\,\Lambda(\vec{x},t) + K\,\sigma^2 s\,\nabla^2 \Lambda(\vec{x},t) - \varepsilon\,\overline{f}\,N(t)\Lambda(\vec{x},t) \qquad (8)$$

The diffusion-controlled species divergence from the centre of radiation can be understood by defining the Fourier transformation of the species density distribution

$$\tilde{\Lambda}(\vec{k};t) = \int_{M_N} d^2 x \exp(i\,\vec{k}\,\vec{x})\,\Lambda(\vec{x}) \qquad (9)$$

Substitution of Eq. (8) in Eq. (9), the decay of each mode is found to be

$$\tilde{\Lambda}(\vec{k};t) \propto \exp[(-k^2\,K\,\sigma^2 s - \varepsilon\,\overline{f}\,N(t) + s)\,t] \qquad (10)$$

By using Eq. (4) to calculate $\tilde{\Lambda}(\vec{k};t)$ [19], the mean square radius of the distribution becomes

$$\left\langle r^2 \right\rangle = \int_{M_N} d^2 x\, x^2\,\Lambda(\vec{x}) \approx 4\sigma^2\,s\,K\,t\,\exp\Big[(s - \varepsilon\overline{f}N(t))\,t\Big] \qquad (11)$$

Here it is assumed that the species density distribution is symmetrical around the radiation centre. For sufficiently large times, the total species number $N(t) \approx s/\varepsilon\overline{f}$ and thus $\left\langle r^2 \right\rangle \approx 4\sigma^2\,s\,K\,t$, hence the process is diffusive with a diffusion constant $D = \sqrt{4\sigma^2 sK}$. However, at the beginning of the radiation, before the species number saturates, the mean square radius of species distribution grows exponentially. One can speculate that the transition between exponential to diffusive growths may parallel the rapid morphological expansion typical of the initial phase of an adaptive radiation followed by slow accretion in the late stages.

## THE BRANCHING PROCESS

Fig. 4A shows an artificial phylogenetic tree obtained by plotting the y-coordinate in $M_N$ of all species present at a given time. The Figure shows that species are not uniformly distributed, rather they appear to form clearly separated lineages. Although species from different lineages compete with each other, lineages do not intersect (the apparent crossing of lineages in Fig. 4A is due to the one-dimensional projection of the plane). By comparison with Fig. 3A, it is clear that the observed step-wise growth corresponds to the creation of new lineages.

Branching is controlled by short-range competition. Consider a radiation stemming from one single species, and let us for simplicity assume a step-wise competition function (model I). In the beginning, species are few and their number grows exponentially. At some point the cluster reaches a local equilibrium density $\Lambda_{eq} = s/(\varepsilon \bar{f})$, where the number of species going extinct per unit time equals the number of speciation events. From this point in time, the cluster will grow in size while maintaining an almost constant number of species. After a time $\tau \approx R^2/(2 s \sigma^2)$, when the diameter of the cluster reaches the size of the short-range competition ($D_{cluster} \approx R$), the competition felt by species at the periphery diminishes because species residing opposite on the cluster becomes sufficiently far. Consequently, peripheral species tend to survive longer and have the potential to originate a new branch. In morphospace one observes that the cluster splits in a characteristic 'fission effect'.

## LINEAGE DYNAMICS

Fig. 4B reports the total number of lineages as a function of time. The graph is made by counting the number of separate clusters at fixed times; despite some uncertainty with the identity of cluster in few cases, the cluster number could be identified with high precision. It is seen that the generation of new lineages is concentrated at the early phase of the radiation.

The hindrance to further branching is an effect of the long-range competition (see Appendix A). With time, when species becomes numerous and occupy larger portions of morphospace, the survival advantage of peripheral species is greatly lost because they begin to experience long-range competition from species belonging to other clusters. Thus, the splitting rate of new lineages drops. Another interesting property of the spatial occupation pattern is the spontaneous appearance of empty zones or voids in morphospace. This effect is a result of increased extinction rate for species occupying the region between approaching lineages.

## DEPENDENCE OF COMPETITION FUNCTION

The influence of the inter-species competition on the temporal occupation pattern in morphospace is studied by simulations using different forms of competition functions (Fig. 5). Provided that the long-range competition remains finite, all cluster separations are found early in the radiation process (Fig. 5A; cf. Fig. 4A). Hence, the results are insensitive to the fine details of the competition function. Fig. 5B is a blow-up of the early phase of the same simulation and shows in more details the branching pattern. Notice the numerous aborted branches, which commonly arise from central species. If the long-range competition goes to zero, the creation of new lineages is found to continue in time (Fig. 5C). This demonstrates the role of long-range competition in dampening the branching at later stages. The morphological clusters tend to occupy regular positions with nearly constant average distance from nearest neighbours (Fig. 6) and branching is visible as "fissioned" clusters.

## *DISCUSSION*

Most of present biodiversity has evolved through explosive diversifications of coexisting species during adaptive radiations [20]. With use of a simple stochastic model we have studied the ecological hypothesis that the ultimate cause of AR's is species competition due to resource depletion, which leads to divergent selection for different habitats and ecological niches.

The present model is aimed at describing the morphological divergence of species in a statistical sense, and the results (Fig. 3-5) bear interesting qualitative resembles with real AR's. First, the morphological diversification of species is faster at the beginning of the radiation and decreases with time, as shown by the temporal mean square radius of the species distribution, which initially growth exponentially and then as a diffusive process. Second, lineages are created spontaneously through branching processes; during further evolution these groups remain isolated from species belonging to other lineages and interact only through long-range competition. Thus, the model suggests that competition is the driving force for creation of voids and patches in morphospace. Third, the creation of new lineages is concentrated in the early stages of the radiation, thereafter their number remains approximately constant with time. Finally, the occupation in morphospace remains limited after the initial phase (although the occupied volume continues to expand slowly) without the need to impose boundaries. In short, the model suggests a scenario for establishment of higher taxonomic groups, which accounts for biological data, without the use of extraordinary evolutionary jumps or phenotypic restrictions. From the numerous simulations performed, we can state that all the significant patterns of the model are qualitatively robust, and independent of the exact form of the competition function.

The conclusions of the present work are restricted as the model does not account for population dynamics but represent each species by a single point. Population-based models are currently investigated in connection with sympatric speciation [21]. The dimensionality of the morphospace affects the maximum number of neighbouring clusters that can exist. In the two-dimensional model explored here, mutually repulsive clusters tend to occupy the plane in a regular lattice at a distance of the order $\sim R$ (neglecting the long-range competition). The maximum number of clusters surrounding a central one is 6, equal to the number of vertices in a regular hexagon. In three dimensions a similar reasoning gives 12 clusters at a distance of $\sim 0.95R$, approximated by the number of vertices in a icosadhedron. Hence, increasing the dimensionality of the morphospace entails greater potential for high-level taxa to be founded. However, the effect is quantitative and does not change the qualitative properties of the model.

Some generalisations of the model have been carried out. (1) Simulations have been performed where each species $S_i$ is assigned with randomly chosen short-range competition ranges $R = R_i$. This modification did not change the general trend of the results. (2) To simulate phyletic evolution, i.e. phenotypic changes in time without real speciation events, a diffusion term was added to the Gaussian distribution of species coordinates in time. The effect of this change was also not noticeable. A more detailed model could introduce predation and trophic levels, although in a morphospace representation it is not conceptually straight-forward how to do this since predation does not necessarily depend on morphological disparity between species.

Finally, it would be interesting to investigate the model predictions on the observed fractal character of taxonomical distribution of organisms.

## APPENDIX:

## ANALYTICAL ESTIMATE OF TEMPORAL BRANCHING RATE

The decreasing branching rate during the progression of an AR is derived analytically. Consider a step-sized competition function given by:

$$f\left(d_{i,j}\right) = \begin{cases} \varepsilon\, f_0 & d_{i,j} < R \\ \varepsilon\, f_\infty & d_{i,j} \geq R \end{cases} \tag{A.1}$$

where $f_0 \gg f_\infty$. With time species become arranged in clusters that develop into separate lineages. Let there be $C$ clusters at a given point in time that each contains $m$ species, where $m \gg 1$. The number of species in a cluster evolves according to

$$\frac{dm}{dt} \approx s\,m - \varepsilon\left[ f_0\,(m-1)^2 + f_\infty\,m^2\,C \right] \simeq s\,m\ - \varepsilon\,m^2\,(f_0 + f_\infty\,C) \tag{A.2}$$

here the first term describes the proliferation of species, the second term is the short-range competition from the $(m-1)$ other species belonging to the same cluster, and the third term is the long-range competition from $m\,C$ species inside external clusters. The equilibrium number of species in a cluster is found from Eq. (A.2) to be

$$\bar{m} = \frac{s}{\varepsilon}\frac{1}{f_0 + f_\infty\,C} \tag{A.3}$$

The species number in a cluster changes slowly during the splitting of a new lineage, and in this case we may use that $m \simeq \bar{m}$.

The generation of new lineages occurs at the edge of clusters where the short-range competition is reduced. The lowered short-range interaction can be quantified by the factor $\chi \leq 1$. Thus, for a central species $\chi \approx 1$, while a species at the cluster border has $\chi < 1$. The probability per unit time that a species will form a new lineage is given by

$$\frac{dp}{dt} = s - \varepsilon \bar{m} \left( f_0 \, \chi + f_\infty \, C \right) = s \left( 1 - \frac{f_0 \, \chi + f_\infty \, C}{f_0 + f_\infty \, C} \right) \qquad \text{(A.4)}$$

where the left hand side expression is found by insertion of Eq. (A.3). The probability has the limiting behaviour

$$\lim_{t \to 0} \left( \frac{dp}{dt} \right) = s \left( 1 - \chi \right) ; \qquad \lim_{t \to \infty} \left( \frac{dp}{dt} \right) = 0 \qquad \text{(A.5)}$$

showing the potential for new lineages to form at early stages when few clusters are present $\left( C \sim 0 \right)$, while in later stages when clusters are numerous, the survival probability of the colonizer drops to zero.

# *FIGURE LEGENDS*

FIG. 1: ADAPTIVE RADIATION OF MAMMALS IN THE TERTIARY

1.a: Total number of mammalian families and orders as function of time; redrawn from [22].

1.b: Animals and plants are classified in a hierarchical taxonomic system according to kingdom, phyla, class, order, family, genera, and species. The figure is redrawn from [23] and shows a graphic representation of evolutionary relationships among orders of mammals (a so-called *phylogenetic tree*). The thickness of each group is proportional to the number of genera in that group.

Notice the rapid establishment of basic lineages (here orders); taxa of lower systematic levels (e.g. families, genera) generally peaks at later times compared to higher orders [24].

FIG. 2: MODEL SCHEMATICS

Species are described by their coordinates in a two-dimensional morphospace defined with reference to two quantifiable phenotypic forms. Species compete with close-by species within a radius of $r < R$ through a short-range competition function and with species further away through a long-range competition function. New species originate as propagules of the parent species within the short-range competition radius.

FIG. 3: SPECIES NUMBER AND MORPHOLOGICAL DISPARITY

3.a: Number of species plotted as function of time; the insert shows the initial phase displaying step-like growth. The simulation is made using a constant short-range competition function with a power-law tail for long-ranges (model II). The constants are: $\sigma = 1$; $R = 20\,\sigma$; $\beta = 0.1$; $\varepsilon = 0.8$; $f_0 = 1$; $s = 100$; $dt = 0.005$; $R_L = 30\,\sigma$.

3.b: The mean square radius of the species density distribution $\langle r^2 \rangle = \sqrt{\left( x_j^2 + y_j^2 \right)}$ plotted as function of time. The competition function and parameters are identical to the case described above.

FIG. 4: PHYLOGENETIC TREE AND LINEAGE DYNAMICS

4.a: A phylogenetic tree constructed by plotting the $y$-coordinate of all living species as function of time.

4.b: The number of lineages plotted as function of time. Note that the morphospace is bi-dimensional, and therefore the occupation patterns are partly masked by the superposition of lineages on the same line of sight.

FIG. 5: LINEAGE DYNAMICS WITH DIFFERENT COMPETITION FUNCTIONS

The number of lineages plotted as function of time; 5.a: Smoothed sphere competition with finite long-range competition (Model I). Parameter values: $\sigma = 1$; $R = 20\,\sigma$; $\varepsilon = 0.8$; $s = 100$; $dt = 0.005$ ; $f_0 = 1$; $f_\infty = 0.2$; a=2. b: Magnification of the first 12.000 time-steps of same simulation; 5.c: Smoothed sphere competition with long-range competition dropping to zero over large distances (Model I). Parameter values as in 5a except that $f_\infty = 0$.

FIG. 6: BRANCHING

Species distribution in morphospace at four different times: Top left: after 1000 time steps from the beginning of the simulation; top right: after 2000 time steps; bottom left: after 3000 time steps; bottom right: after 8000 time steps. The figure shows the dynamics of clusters "fission" during branching and the regular cluster distribution in morphospace. Model II with $\sigma=1$; $R=18\,\sigma$; $\beta=0.1$; $\varepsilon=0.8$; $f_0=1$; $s=80$; $dt=0.005$; $R_L=30\,\sigma$.

[24] The numbers between 1-21 refer to extinct orders and scattered genera usually poorly represented in the fossil record. 22: Marsupialia, 23: Edentata, 24: Lagomorpha, 25: Rodentia, 26: Primates, 27: Dermoptera, 28: Chiroptera, 29: Insectivora, 30: Carnivora, 31: Cetacea, 32: Artiodactyla, 33: Tubulidentata, 34: Perissodactyla, 35: Hyracoidea, 36: Proboscidea, 37: Sirenia, 38: Monotremata.

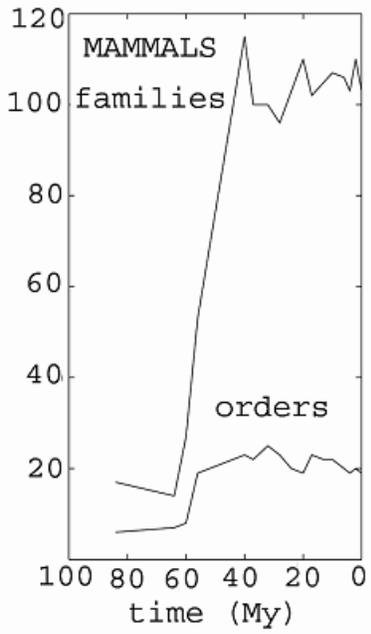

figure 1a

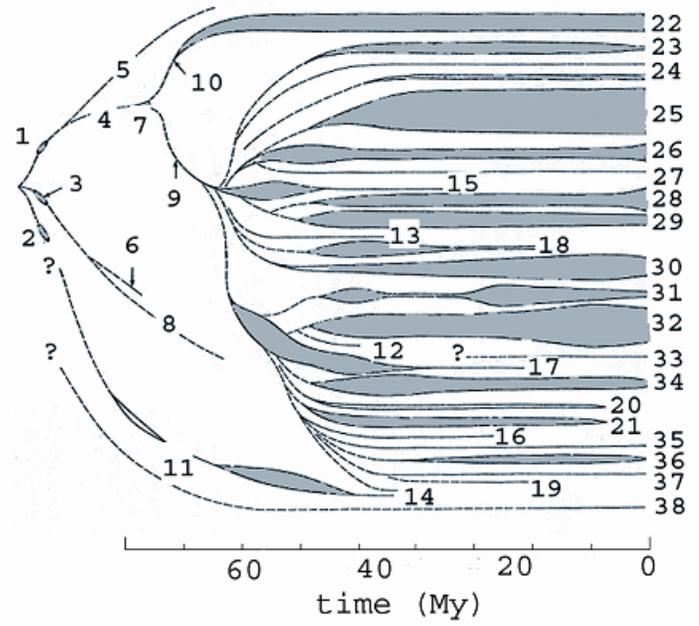

figure 1b

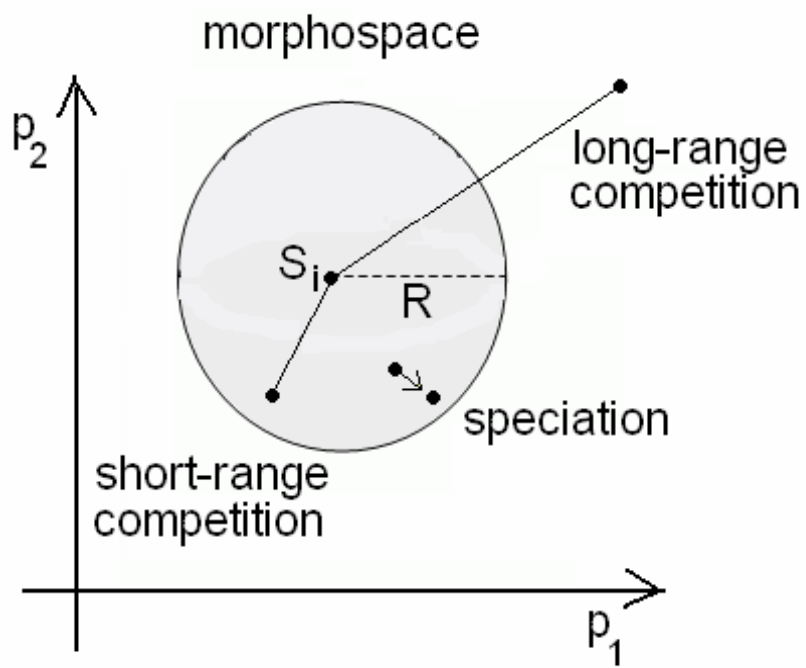

figure 2

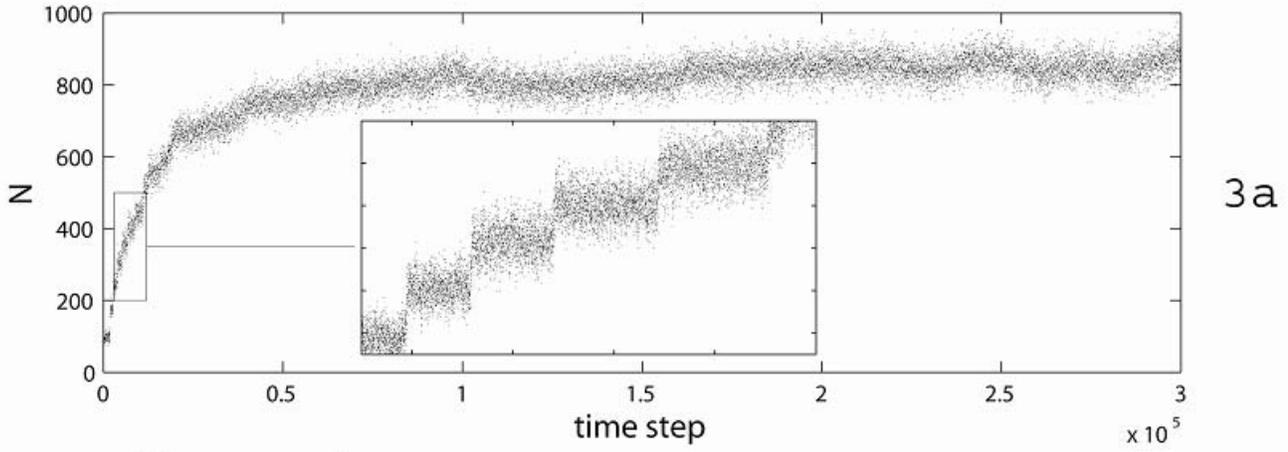

figure 3a

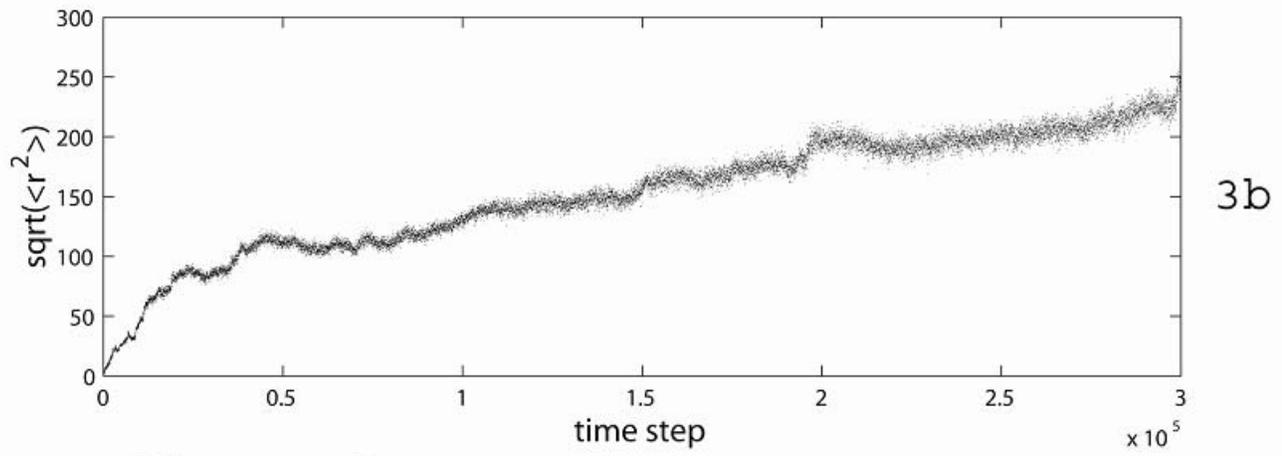

figure 3b

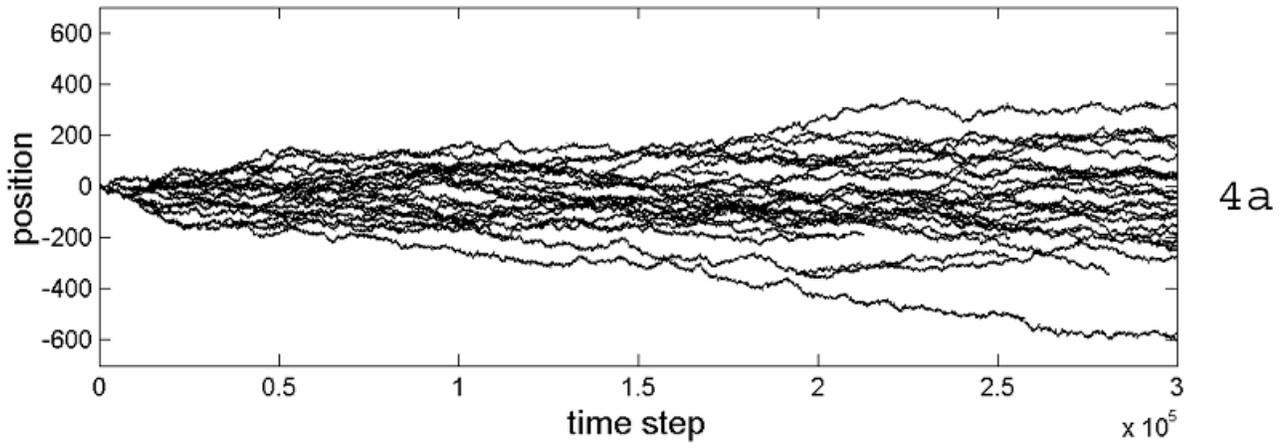

4a

figure 4a

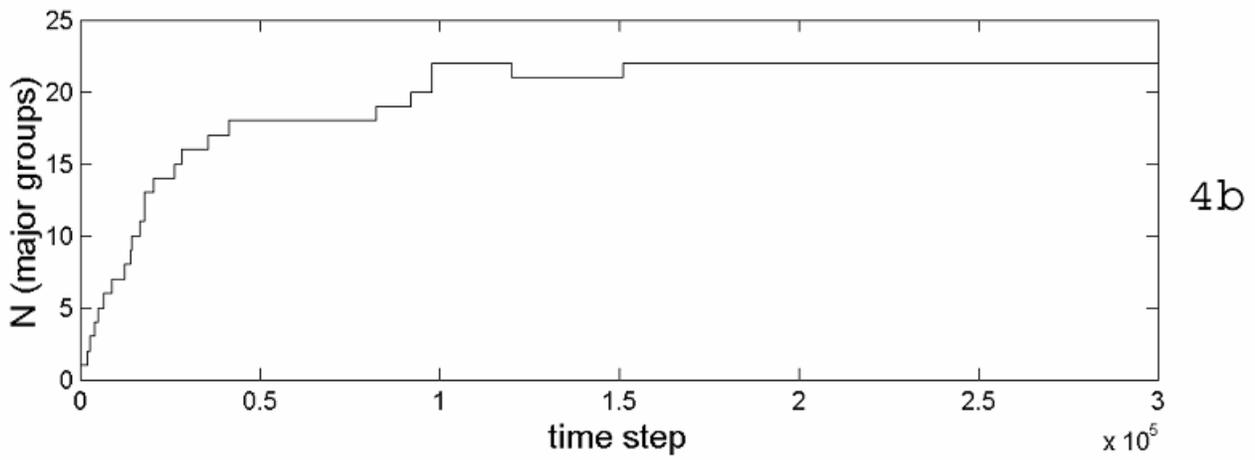

4b

figure 4b

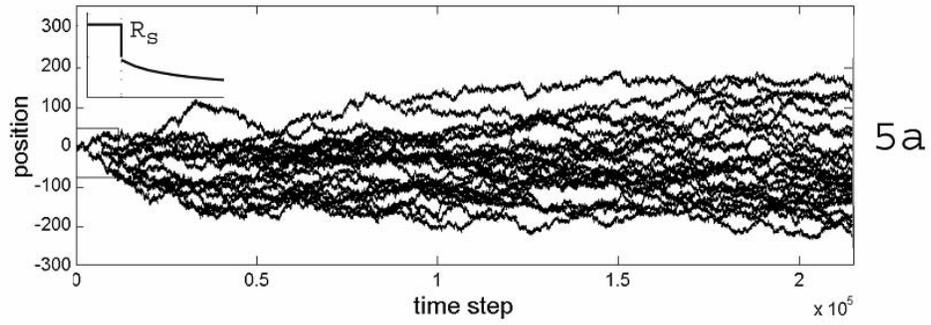

5a

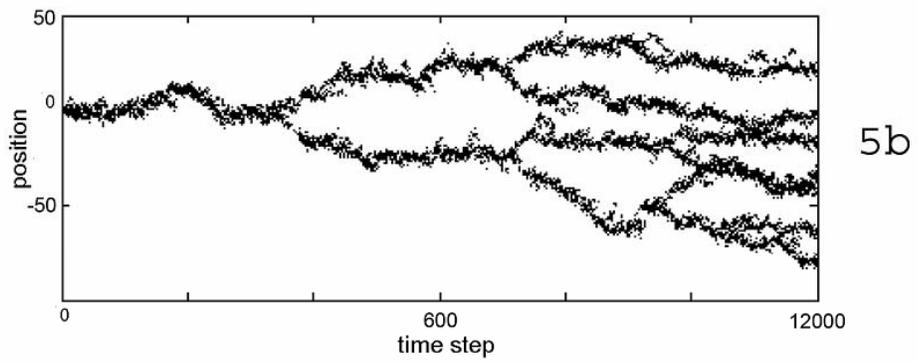

5b

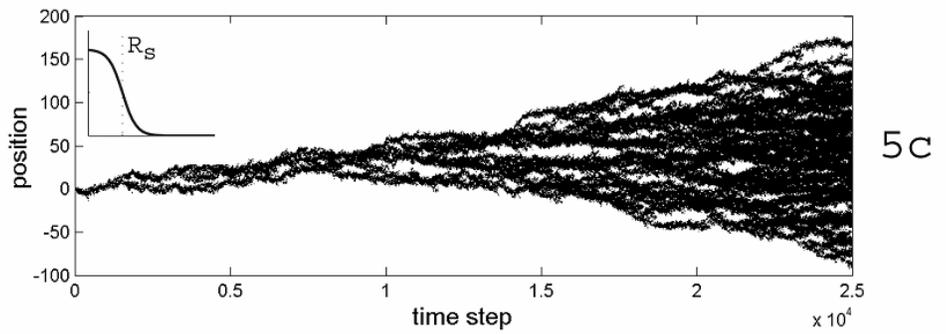

5c

figure 5

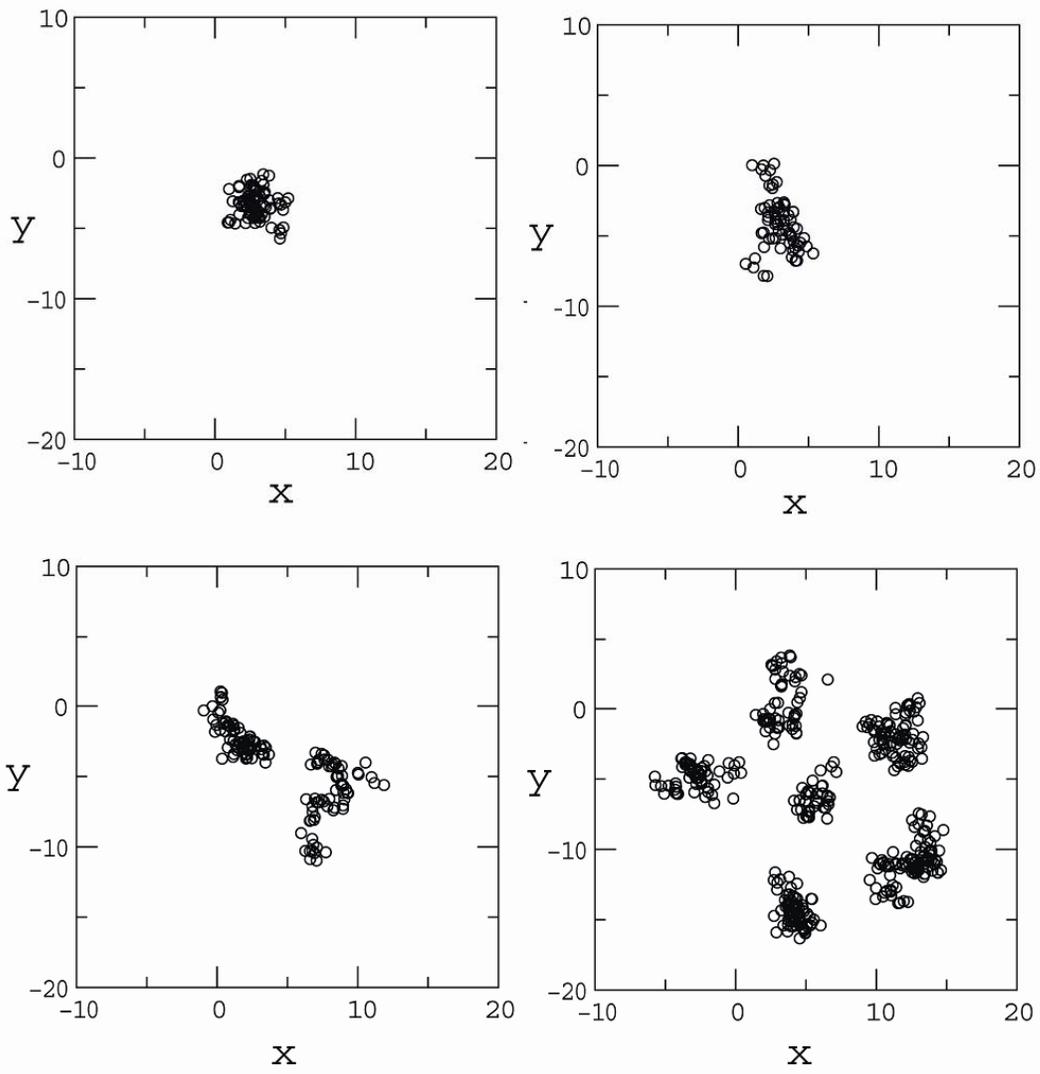

figure 6